\documentclass[10pt]{iopart}
\usepackage{graphicx}
\usepackage{bm}

\usepackage[utf8]{inputenc}
\usepackage[T1]{fontenc}
\usepackage{mathptmx}
\usepackage{etoolbox}
\usepackage[normalem]{ulem}
\usepackage{color}
\makeatletter
\def\@email#1#2{%
 \endgroup
 \patchcmd{\titleblock@produce}
  {\frontmatter@RRAPformat}
  {\frontmatter@RRAPformat{\produce@RRAP{*#1\href{mailto:#2}{#2}}}\frontmatter@RRAPformat}
  {}{}
}%
\makeatother
\begin{document}

\title{Enhancement of the Magnetocaloric Effect in Geometrically Frustrated Cluster Spin Glass Systems
}
\author{F. M. Zimmer$^{1,a}$, R. Mourão$^1$, M. Schmidt$^{2}$, M. A. Tumelero$^{3}$ and S. G. Magalhaes$^{3}$, }
\ead{$^a$fabiozimmer@gmail.com}
\address{$^1$ Instituto de F\'isica, Universidade Federal de Mato Grosso do Sul, Campo Grande-MS, Brazil \\
$^2$ Departamento de F\'isica, Universidade Federal de Santa Maria, 97105-900, Santa
Maria, RS, Brazil \\ 
$^3$ Instituto de F\'isica, Universidade Federal do Rio Grande do Sul,  91501-970, Porto Alegre, RS, Brazil  }

\date{\today}

\begin{abstract}

In this work, we theoretically demonstrate that a strong enhancement of the Magnetocaloric Effect is achieved in geometrically frustrated cluster spin-glass systems just above the freezing temperature.  We consider a network of  clusters interacting randomly which have
triangular structure composed  of Ising spins interacting antiferromagnetically. 
The intercluster disorder problem is treated using a cluster spin glass mean-field theory, which allows exact solution of the disordered problem. The intracluster part can be solved using exact enumeration. 
The coupling between the inter and  intracluster problem incorporates the interplay between effects coming from geometric frustration and disorder. As a result, it is shown that there is the onset of cluster spin glass phase even with very weak disorder.
Remarkably, it is exactly within a range of very weak disorder and small magnetic field that is observed the strongest isothermal release of entropy. 
\end{abstract}

\maketitle
\ioptwocol

\section{\label{sec:intro}Introduction}
Frustrated magnetism has been an exciting subject 
for searching new states and new properties of matter \cite{Lacroix_book, ramirez}. 
The immense manifold of degenerated spin configurations of magnetic frustrated systems reflects the definitive  impossibility of spins to simultaneously satisfy the interactions to which they are subjected.
This huge spin degeneracy ties frustrated magnetism and magnetic refrigeration, i.e., cooling or heating achieved by adiabatic variations of magnetic field, since it provides a suitable mechanism for the enhancement of the  magnetocaloric effect (MCE) \cite{doi:10.1146/annurev-matsci-062910-100356,doi:10.1063/1.4983612}.
The technological applications of MCE have been recognised as a potential substitute for the environmentally harmful conventional refrigeration technologies in a scenario where 
the search for sustainable development is imperative to prevent climate change and, therefore, it is necessary to achieve an increasingly improvement of energy efficiency 
\cite{Franco2018}. 

The sources of frustration can come from competing nearest neighbor and further neighbor interactions, special lattice geometries with only nearest neighbor interaction and the combination of disorder and competing ferromagnetic and antiferromagnetic interactions.  The two former cases can evade magnetic order. The latter case can give raise to the spin glass physics with its non-trivial broken ergodicity \cite{Mezard87}.  For  non-disordered cases, 
there are theoretical predictions indicating that the enhancement of the MCE 
can be obtained in kagome,  pyrochlore and garnet lattices \cite{PhysRevB.67.104421} and in the $J_1$-$J_2$ square lattice \cite{PhysRevB.76.125113} with both cases having Heisenberg spins as well as the pyrochlore lattice with Ising spins \cite{PhysRevE.96.052128}. These predictions have been confirmed for Gd$_2$Ti$_2$O$_7$ \cite{PhysRevB.71.094413}.  

Interestingly, there are also recent observations of enhanced MCE in spin glass-like systems such as R$_2$NiSi$_3$ (R=Gd,Er) \cite{PhysRevB.94.104414},  NdNi$_{0.94}$Si$_{2.94}$ \cite{Pakhira2018}, Pr$_2$Ni$_{0.95}$Si$_{2.95}$ \cite{Pakhira2019}, Tb$_2$Ni$_{0.94}$Si$_{3.2}$, GdCu$_4$Mn \cite{Karol2022}, Ho$_2$CoMnO$_6$ \cite{Patra2022} indicating that this manifestation of frustration due to disorder and competing interactions can be also explored for magnetic refrigeration. In addition, cluster-glass systems of some binary intermetallic compounds R$_5$Pd$_2$ (R=rare-earth ions) \cite{Gubkin2013} 
have also reported giant MCE 
\cite{Sharma2018, Sharma2018a, Saori2015, Paramanik2015} as well as Dy$_5$Pd$_{2-x}$Ni$_x$ \cite{Mohit2019}.
Other MCE systems of TbLaSiGe have shown a novel cluster glass state within a Griffiths phase appearing above the ferromagnetic ordering \cite{PhysRevB.99.054419}. These are part of a strong group of materials showing giant MCE. 

In view of evidences that frustration produced by different sources can be effective in enhancing MCE, one natural next step is to ask how would be the effects of combined sources of frustration on the MCE. 
In other words, could one expect an even greater amount of magnetic entropy released in the polarization process of, for instance, 
geometrically frustrated systems 
with a stabilized 
spin glass-like phase at lower temperature?

    The question above is very intertwined with 
    the interplay between disorder and geometric frustration. Indeed, this is an open question that goes through understanding whether geometric frustration (GF) can stabilize a spin glass-like state \cite{RevModPhys.82.53}. To address this  challenging problem, the theory must necessarily overcome or deeply modify prototypical spin glass (SG) theories as, for instance,  the mean field replica symmetry Sherrington-Kirkpatrick model \cite{PhysRevLett.35.1792} and the Parisi's replica symmetry breaking approach \cite{Parisi_19802, Parisi_19803}.  This has been the subject of several interesting new proposals for the role of disorder in  geometrically frustrated SG systems \cite{PhysRevB.81.014406, PhysRevLett.110.017203} or even to
    account for the glassy behavior  
    in the absence of disorder \cite{PhysRevB.47.15342}.

    A particularly interesting approach to treat  SG systems  with GF 
    is the description of the problem in terms of spin clusters.  Among other advantages, this approach also allows a direct connection between GF, disorder, molecular nanomagnetism
    \cite{Schnack2010} and MCE \cite{Nair2018}. 
    One way to achieve that is through
    the cluster spin glass 
        mean field  
    theory \cite{Soukoulis78-1,Soukoulis78-2}.
    Recently, this theory has been used to investigate the interplay between disorder and GF within the standard mean field approximation \cite{PRE_Zimmer2014, PRE_Zimmer2014_2, PhysA_Schmidt2015, JPCM_Schmidt2017} and also employing random graphs \cite{PhysRevE.103.052110}. 
    Basically, in the  cluster spin glass 
    mean field  theory, the original disordered spin lattice model 
    is clusterized, i. e., it 
    is splitted into a long range random intercluster 
    (which stabilizes a spin glass-like phase) 
    and a short range intracluster  interactions. Within the mean field approximation, the stabilization of a spin glass-like phase comes from the solution of self-consistent equations of two free energy derived variational  parameters: 
    the spin glass order parameter 
     and the
   cluster magnetic moment self-interaction $\bar{q}$.

    It is important to remark that, for the purposes of this work, 
    the temperature range of interest for the MCE is within  the paramagnetic phase.  We highlight that it is the frustration inside the spin clusters  which is the cornerstone for the enhancement of the MCE  when the magnetic field is turned on. 
    Indeed, geometrically  frustrated isolated Ising spin clusters have been proposed already to increase MCE  \cite{ZUKOVIC201522}. 
    Nevertheless, our effective cluster model refers, in fact, to a disordered network of clusters which is treated at mean field level. 
    Ultimately, it is from the delicate balance between the GF effects (coming from the spin cluster)  and the disorder (related to the network of spin clusters) in the presence of magnetic field, that we shall obtain 
    enhancement of the MCE.

   The paper is organized in the following sequence: the model and theoretical framework are presented in section II, numerical results and discussions on the MCE are presented in section III, conclusion and further remarks in the last section.

\section{
Model, and Replica Procedure}
\label{sec:method}

We consider a system divided into clusters  with two kinds of interactions able to introduce frustration: the intracluster one that regards spins on triangular clusters with antiferromagnetic interactions, and long-range quenched disordered interactions between clusters. The model is written as
\begin{equation}\label{hamiltonian}
    H=-\sum_{\nu,\lambda} J_{\nu\lambda} \sigma_{\nu}\sigma_{\lambda}+\sum_{\nu}^{N_{cl}}(\sum_{(i,j)} J_1 \sigma_{\nu_i}\sigma_{\nu_j}-h\sum_{i}^{n_s} \sigma_{\nu_i}),
\end{equation}
where $\sigma_{\nu}=\sum_{i}^{n_s} \sigma_{\nu_i}$ corresponds to the magnetic moment of the cluster $\nu$, with $\sigma_{\nu_i}$ representing  the Ising spin at the site $i$ of cluster $\nu$,  $n_s$ is the number of sites per cluster and $N_{cl}$ represents the number of clusters. 
The interactions $J_{\nu\lambda}$ between cluster magnetic moments are random variables that follow gaussian probability distributions with average zero and standard deviation $J/\sqrt{z_c}$, in which $z_c$ is the number of neighbor clusters \cite{Soukoulis78-1,Soukoulis78-2}. The antiferromagnetic intracluster interactions ($J_1>0$) consider only nearest neighbors spins in clusters with triangular lattice structures \cite{PRE_Zimmer2014} and $h$ stands for the external magnetic field.

The problem is treated with the replica method \cite{PhysRevLett.35.1792} employed to evaluate the intercluster disordered interactions. This allows to write the free energy per cluster as $f=-\lim_{n\rightarrow 0} (Z(n)-1)/(N_{cl} n \beta)$, where the replicated partition function becomes
\begin{equation}
    Z(n)=\mbox{Tr}~ \exp{\left[\frac{\beta^2 J^2}{2z_c}\sum_{\nu,\lambda}\sum_{\alpha,\gamma} \sigma_{\nu}^{\alpha}\sigma_{\lambda}^{\alpha}\sigma_{\nu}^{\gamma}\sigma_{\lambda}^{\gamma}-\beta\sum_{\alpha}\sum_{\nu}H_{intra}^{\nu,\alpha}\right]}\label{replicated}
\end{equation}
where Tr is the trace over spin variables,  $\alpha$ ($\gamma$) represents the replica index, $\beta=1/{k_B T}$  and $H_{intra}^{\nu,\alpha}$ corresponds to the two last terms of Eq. (\ref{hamiltonian}) with an additional replica index $\alpha$ in the spin variables.

The intercluster terms of Eq. (\ref{replicated}) are decoupled by introducing variational mean-field parameters  $q^{\alpha\gamma}$ \cite{Soukoulis78-2}, 
resulting in the following free energy:
\begin{equation}
    f=\frac{\beta J^2}{4n}\sum_{\alpha,\gamma}(q^{\alpha\gamma})^2-\frac{1}{\beta n}\ln \mbox{Tr}~ \exp{[-\beta H_{eff}]},
\end{equation}
with the effective replica-dependent single-cluster model 
$H_{eff}=-\beta J^2/2\sum_{\alpha\gamma} q^{\alpha\gamma}\sigma^{\alpha}\sigma^{\gamma}-\sum_{\alpha}H_{intra}^{\alpha}$.
In particular, the introduced variational parameters follow self-consistent equations that represent the cluster magnetic moment self-interaction $q^{\alpha\alpha}=\langle \sigma^{\alpha}\sigma^{\alpha}\rangle_{H_{eff}}$ and the cluster spin-glass order parameter 
$q^{\alpha\gamma}=\langle \sigma^{\alpha}\sigma^{\gamma}\rangle_{H_{eff}}$ (when $\alpha\neq\gamma$), where $\langle\cdots \rangle_{H_{eff}}$ stands for the thermal average over the effective model $H_{eff}$ \cite{PRE_Zimmer2014,JPCM_Schmidt2017}.

As follows, we assume the replica symmetry  solution ($q=q^{\alpha\gamma}$ and $\bar{q}=q^{\alpha\alpha}$) and apply a Hubbard-Stratonovich transformation to decouple the term $(\sum_{\alpha} \sigma^{\alpha})^2$. This procedure allows obtaining the free energy of a single-cluster as: 
\begin{equation}\label{free}
    f=\frac{\beta J^2}{4}(\bar{q}^2-q^2)-\frac{1}{\beta}\int Dx \ln \mbox{Tr} \exp{[-\beta H_{eff}(x)]}
\end{equation}
where 
\begin{equation}
H_{eff}(x)=J_1\sum_{(i,j)}\sigma_i\sigma_j-J\sqrt{q}z\sigma-\frac{\beta J^2(\bar{q}-q)}{2}\sigma^2-h\sum_{i}^{n_s}\sigma_i
\end{equation}
with $Dx=dx\exp{(-x^2/2)/\sqrt{2\pi}}$ and the order parameter $q$ and $\bar{q}$ are chosen to extremize the free energy (\ref{free}): $\frac{\partial f}{\partial q}=0$ and $\frac{\partial f}{\partial \bar{q}}=0$. In other words, the effective model preserves all intracluster dynamics, which can be evaluated exactly by exact diagonalization (or exact enumeration), while the intercluster interactions enter as self-consistent fields given by $q$ and $\bar{q}$. Other thermodynamics quantities as magnetization $m$ and magnetic entropy $S$ per site can be derived from the free energy:  $m=-\frac{\partial f}{\partial h}/n_s=\int Dx\langle \sum_{i}\sigma_i\rangle_{H_{eff}(x)}/n_s$ and $S=-\frac{\partial f}{\partial T}/n_s$. The access to the system entropy allows the straightforward evaluation of the isothermal entropy change $\Delta S = S(T,h)-S(T,0)$, which is of utmost importance for the characterization of magnetocaloric properties.  The stability of the replica symmetry  solution is analyzed by the de Almeida-Thouless eigenvalue (replicon) \cite{Almeida_1978}:
\begin{equation}
 \lambda_{AT}=\beta^2 J^2 - \beta^4 J^4 \int Dx [\langle \sigma\rangle_{H_{eff}(x)}^2-\langle \sigma^2\rangle_{H_{eff}(x)}]^2.   \label{at}
\end{equation}

\section{Results}
 In this Section, we analyze our numerical results for the thermodynamics  of the effective cluster model, building a picture for the magnetocaloric potential at different levels of disorder in geometrically frustrated systems. The role of geometric frustration is incorporated by adopting triangular lattice clusters of different sizes.
The stability of the replica symmetry  solution is evaluated from the de Almeida-Thouless eigenvalue $\lambda_{AT}$ (eq. (\ref{at})). The cluster spin-glass (CSG) solution is found when $\lambda_{AT}<0$, while the paramagnetic (PM) phase takes place for a positive eigenvalue $\lambda_{AT}$.
For numerical purposes, we adopt $k_B=1$ and consider $J_1$ as an energy scale. In this case, we remark that $k_B$ becomes implicit in the entropy unit ($S/k_B\rightarrow S$) and in the product $k_B T$ ($k_B T\rightarrow T$). In the following, we discuss the clean limit of the model ($J=0$), providing  a baseline for the discussion of disorder effects, which will be given in Section \ref{disorder}.

\begin{figure}[!ht]
\includegraphics[width=0.95\linewidth]{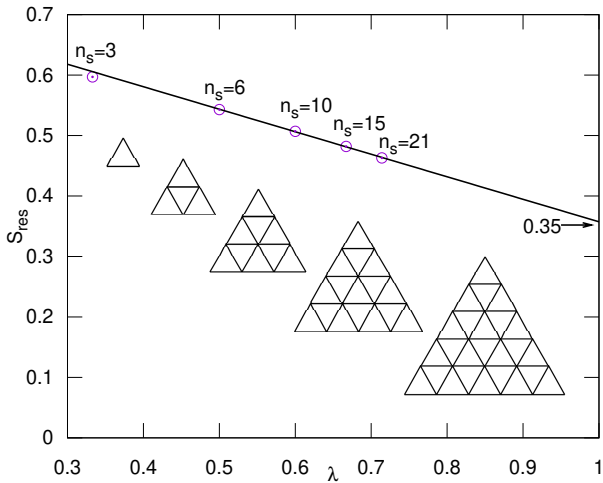}
\caption{Scale theory for the residual entropy $S_{res}$,  adopting isolated-finite clusters in the clean disorder limit. The scaling parameter $\lambda= 2 n_b/(n_s z)$, where $n_b$ is the number of bonds within the cluster and $z$ is the coordination number ($z=6$ for the triangular lattice). The line represents the linear extrapolation of the residual entropy for several cluster size, in which $\lambda=1$ corresponds to the estimation for the triangular lattice result in the thermodynamic limit \cite{PhysRevB.86.054516}. }\label{clean_limit}
\end{figure}
\subsection{The disorder-free limit}

Size and shape of finite clusters can affect thermodynamic quantities, specially in highly frustrated systems as the triangular lattice \cite{ZUKOVIC201522}.  However, a finite size analysis is expected to provide an accurate estimate for the thermodynamic limit. In the present work, we consider triangular shaped clusters of  $n_s$ sites on a triangular lattice, as shown in the detail of Fig. \ref{clean_limit}. In the clean limit ($J=0$), these decoupled clusters incorporate the macroscopic degeneracy characteristic of the triangular lattice, which is manifested in the residual entropy. For instance, Fig.  (\ref{clean_limit}) exhibits a scale analysis for the residual entropy per site of the cluster sizes and shapes adopted in the present work. Here, $\lambda= 2 n_b/(n_s z)$, where $n_b$ is the number of bonds within the cluster and $z$ is the lattice coordination number. The line represents the linear extrapolation of the residual entropy of the clusters considered, in which the thermodynamic limit is expected to be recovered at $\lambda=1$  \cite{PhysRevB.86.054516}. It is worth noting, that the linear fit yields a residual entropy per spin of 0.35 (in units of $k_B$), which is remarkably close to the exact solution for the lattice (0.32) \cite{PhysRev.79.357, PhysRevB.7.5017}. It means that the adopted cluster shapes preserve features of the triangular lattice relevant for the entropic content of this highly frustrated system. 

\begin{figure}[!t]
\includegraphics[width=0.95\linewidth]{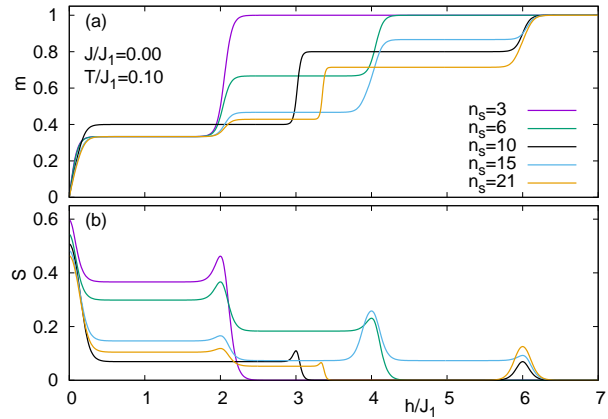}
\caption{
(a) Magnetization $m$ and (b) entropy $S$ versus the external magnetic field $h$ for several clusters size $n_s$ at a finite temperature $T/J_1=0.10$ in the clean quenched disorder limit ($J/J_1=0$). The magnetization shows a plateaus structure that depends on the $n_s$.  In panel (b), an entropy release takes place when the magnetization jump between plateaus occurs with increasing of the field. However, close to these critical fields, the entropy presents a small increasing with magnetic field before it be released. This entropy behavior is also dependent on the cluster size.}\label{entropy_clean}
\end{figure}

A longitudinal magnetic field can drive interesting effects on frustrated finite cluster systems, such as magnetization and entropy plateaus \cite{ZUKOVIC201522, magnetochemistry6040056, STRECKA201576, ZUKOVIC2018311, MOHYLNA20192525}. In Fig. \ref{entropy_clean}, we present the low-temperature behavior of magnetization and entropy, in which signatures of the ground-state plateaus can be spotted.
Magnetization jumps can be found at critical fields as $h/J_1 = 2, ~3, ~3.5, ~4$ and 6, depending on the cluster size. The degeneracy at the critical field drives a higher entropy between plateaus. In other words, the magnetization "jumps" are accompanied by an entropy increase, before it drops at the following plateau, which  is an interesting feature to explore in the MCE \cite{ZUKOVIC201522}. 

We note that a significant entropy drop is driven by a small magnetic field. Moreover, the cluster size increasing leads to a reduction in the entropy of this small field plateau. This entropy drop for small intensities of magnetic field can be interesting to explore the magnetocaloric potential of geometrically frustrated cluster systems. However, the $n_s=10$ cluster shows a more significant reduction of entropy, deviating from the tendency observed for other $n_s$, which might be related to the cluster size not being a multiple of three. The cluster size also prevents a 1/3 magnetization plateau and this cluster is the only one that does not show a magnetization jump at $h/J_1=2$. It is worth to remark that this 1/3 plateau has been reported for the infinite  Ising triangular lattice \cite{Hu_2008}. Therefore,  we neglect the $n_s=10$ cluster from our analysis of disorder effects, which is done in the following section.

\begin{figure}[!t]
\includegraphics[width=0.98\linewidth]{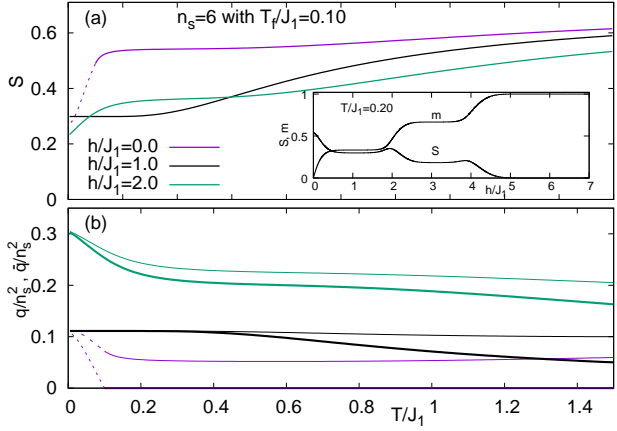}
\caption{(a) Entropy and (b) CSG order parameter (thick lines) and the cluster magnetic
moment self-interaction (thin lines) as a function of the temperature for different external magnetic fields at low intensity of disorder $J/J_1=0.0385$ for clusters with six sites, $n_s=6$
(see cluster shape in Fig. \ref{clean_limit}  and the relation between $T_f/J_1$ and $J/J_1$ is presented in Fig. \ref{tf_vs_j}). The dashed lines indicate the CSG phase with unstable replica symmetry  solution ($\lambda_{AT}<0$). The inset exhibits the magnetization and entropy as a function of the field for a constant temperature $T/J_1=0.2$. }
\label{q_qb}
\end{figure}

\begin{figure}[!t]
\includegraphics[width=0.98\linewidth]{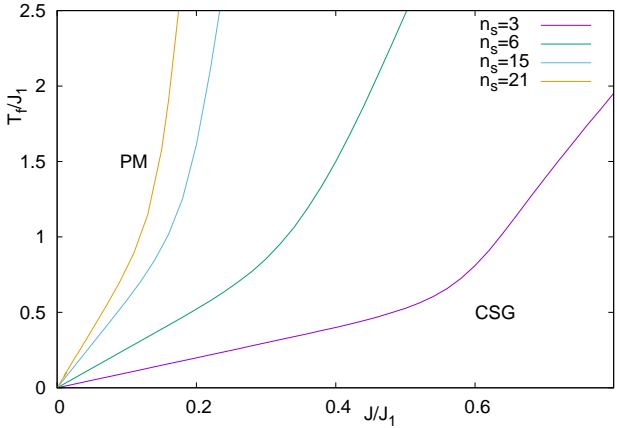}
\caption{The temperature versus disorder strength phase diagrams for systems composed by triangular clusters ($n_s$ = 3, 6, 15 and 21) with AF interactions ($J_1$) on a network with disordered inter-cluster interactions. The clusters shapes are the same as in Fig. \ref{clean_limit}.}
\label{tf_vs_j}
\end{figure}

\begin{figure*}[!t]
\includegraphics[width=0.99\linewidth]{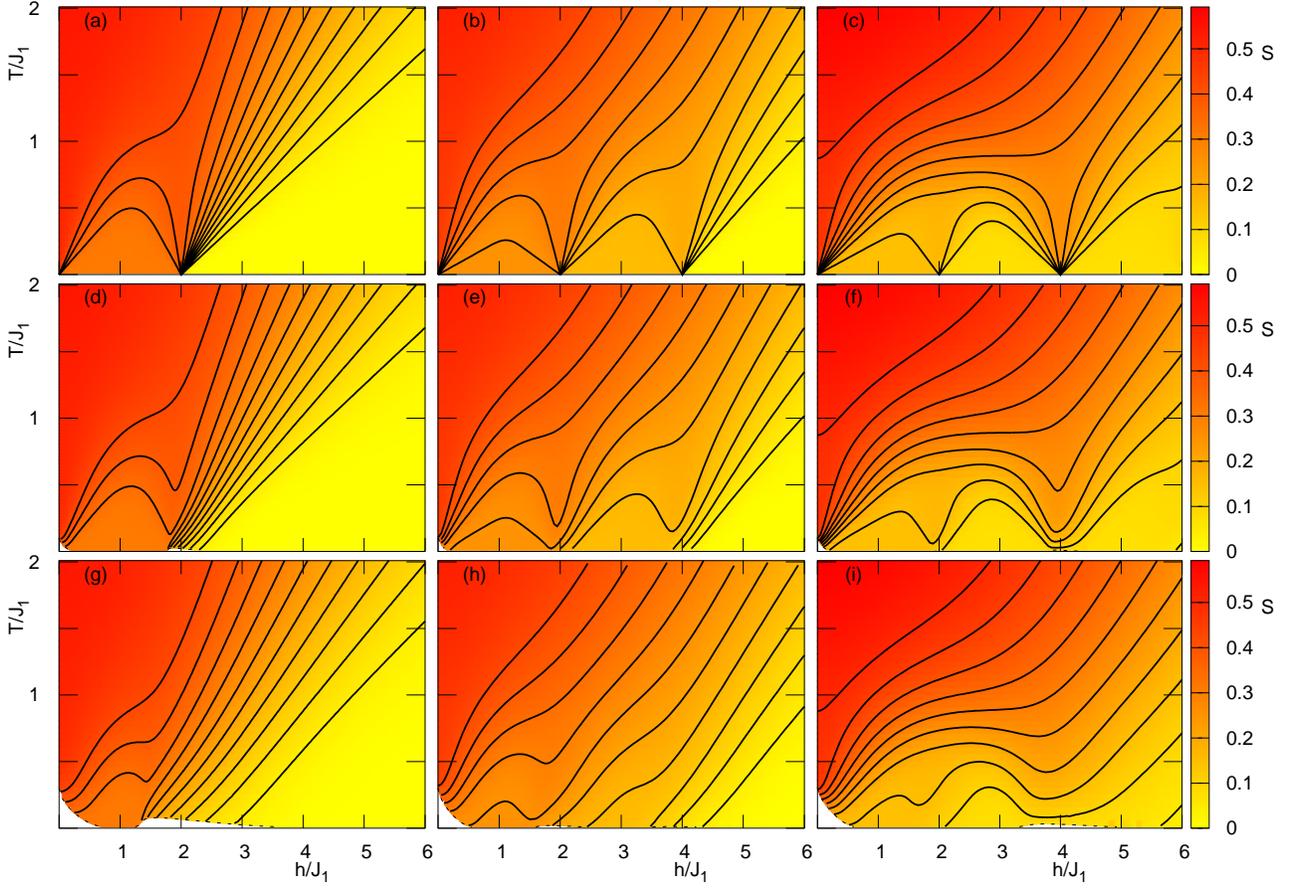}
\caption{Entropy $S$ as a function of $h/J_1$ and $T/J_1$, in which the curves represent isentropes ($S=0.50, \ 0.45, \cdots 0.05$). From the left to right panels are for clusters with 3, 6 and 15 sites, respectively, in which the clean limit (panels (a)-(c)), weak (panels (d)-(f)) and intermediate (panels (g)-(i)) disorders are presented with  $T_f/J_1(h=0)=0.00$ 0.10 and 0.30, respectively.  }
\label{iso_entropic}
\end{figure*}

\begin{figure*}[!t]
\includegraphics[width=0.95\linewidth]{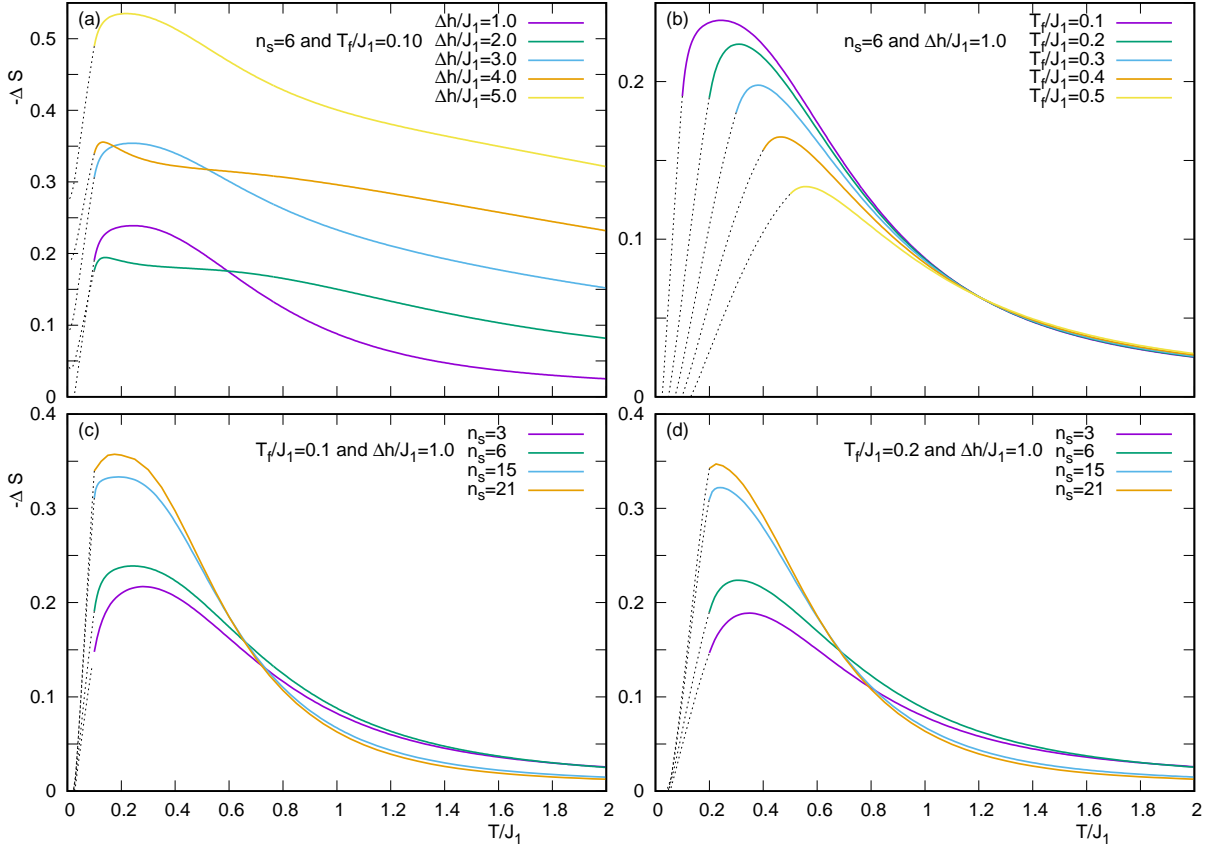}
\caption{Isothermal entropy change as a function of the temperature for several disordered clusters configurations. Panels (a) and (b) consider clusters with 6 sites, in which the intensity of external magnetic field variation $h/J_1$ are changed for $T_f/J_1=0.1$ in (a) and the disorder strength is increased for a constant $h/J_1=1.0$ in panel (b). Panels (c) and (d) exhibit the isothermal entropy change for different cluster sizes in a weak external field ($h/J_1=1.0$) with disorder adjusted to leads to $T_f/J_1=0.1$  and 0.2, respectively. The dotted lines stand for  replica symmetry breaking regime. }
\label{delta_s}
\end{figure*}

\begin{figure*}[!t]
\includegraphics[width=0.95\linewidth]{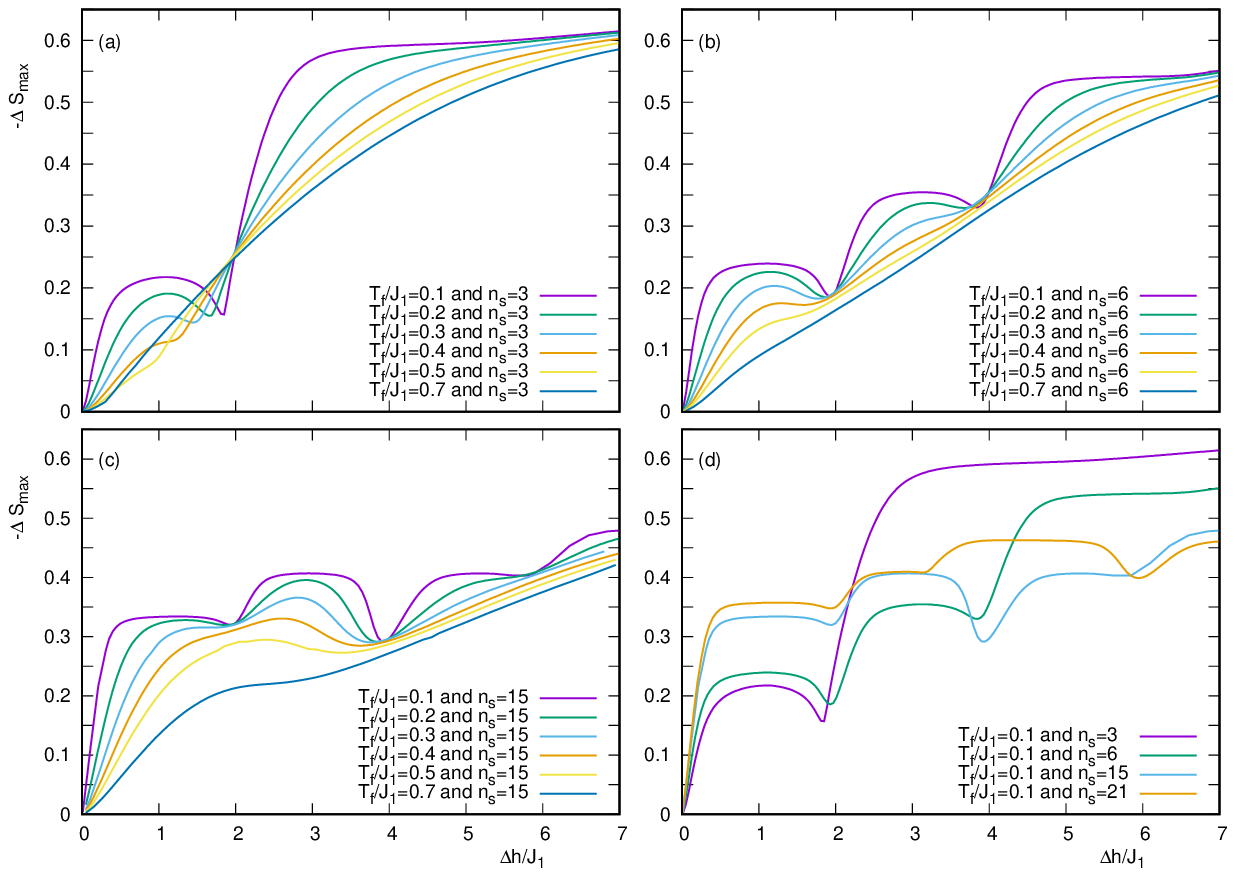}
\caption{Maximum value of the isothermal entropy variation, $-\Delta S_{max}$, as a function of $\Delta h$ for different disorder strengths and cluster size. Panels (a), (b) and (c) are for $n_s=3,~6$ and 15, respectively. Panel (d) exhibits a comparison of $-\Delta S_{max}$ for different cluster sizes at weak disorders.
}
\label{max_entropy}
\end{figure*}

\subsection{The role of disorder on magnetocaloric effect}\label{disorder}

We start our discussion of the role of disorder ($J/J_1>0$) on geometrically frustrated clusters by analyzing the behavior of the entropy and parameters $q$ and $\bar{q}$ for clusters with six sites (see Fig. \ref{q_qb}).  In the absence of external magnetic field, $S$ exhibits a high value in a certain range of temperature slightly above the CSG phase and then drops when the CSG occurs. In particular, the freezing temperature $T_f$ is the borderline between the solutions with  replica symmetry stable ($\lambda_{AT}>0$) and  unstable ($\lambda_{AT}<0$, with dotted lines in Fig. \ref{q_qb}), where $q>0$ for the case $h/J_1=0$. When the external field is turned on, the entropy decreases as compared with the case $h=0$. This entropy difference caused by the field for a given temperature is used to obtain the entropy variation, which is discussed in detail below. Furthermore, the order parameter $q$ becomes induced by the field $h$ in the whole temperature range (see Fig. \ref{q_qb}-b), and it can not be adopted anymore to locate the $T_f$. Instead of $q$, the replicon eigenvalue $\lambda_{AT}=0$ locates the $T_f$ in the presence of magnetic fields.
The cluster magnetic moment self-interaction $\bar{q}$ is also affected by the magnetic field that can polarize spins of the cluster, increasing  the cluster magnetic moment.  It is important to remark that the behavior of $\bar{q}$ reflects the cluster spin states, which are also temperature dependent. In particular, at the highest temperatures, all cluster spin states become equiprobable, leading $\bar{q}\rightarrow n_s$. However, when the temperature decreases, some of the spin states become energetically favorable. For AF intracluster interactions in the absence of external field, the states with low cluster magnetic moment are energetically favorable, driving to lower values of $\bar{q}$ \cite{PRE_Zimmer2014}. In addition, the magnetization as a function of the field presents plateaus as a consequence of the geometric frustration. The switch between these plateaus is followed by an entropy release (see inset of Fig. \ref{q_qb}-a)

In order to  follow 
the disorder effects on the geometrically frustrated clusters, it  is important to analyze possible phase diagrams of the system  at zero magnetic field.
Figure \ref{tf_vs_j} shows the  freezing temperature $T_f$ as a function of the strength of disorder $J$ for several cluster size in  the absence of external magnetic field. The CSG/PM phase boundary is located by the de Almeida-Thouless line ($\lambda_{AT}=0$ in Eq. (\ref{at})).
These phase diagrams exhibit two  disordered regimes, as indicated by the two different inclinations of the $T_f/J_1$ critical line;
one for lower disorder regime, where GF effects are relevant; another for higher disorder regime, in which the disordered couplings are dominant (higher slope of $T_f(J)$). 
The crossover between these regimes depends on the cluster size. Although the intensity of  the disorder needed to reach the conventional CSG regime decreases as the cluster size increases, the freezing temperature becomes slightly higher for larger clusters. 
It means that the 
cluster size increase allows GF effects at higher temperatures on the present cluster model. In other words, larger clusters account for more intracluster frustrated interactions, which are exactly considered. 
It is worth stressing that both disorder strength and cluster size rule the energy scale, playing a crucial role in the freezing temperature.
Therefore, in the following, we characterize regimes of weak and strong disorder considering the effects of the combination of $J/J_1$ and cluster size. 
In this sense, the comparison of different clusters size results is done by adjusting the disorder intensity $J/J_1$ to get the same freezing temperature at zero magnetic field.

The role 
of temperature and external magnetic field on the entropic content of this disordered model is displayed in Fig. \ref{iso_entropic}.
In particular, it exhibits  results for clusters with 3 (first column), 6 (second column) and 15 (third column) spins 
in different regimes of intercluster disorder: clean limit (upper panels), weak (middle panels  with $T_f=0.1 J_1$ at $h=0$) and intermediate (bottom panels with $T_f=0.3 J_1$ at $h=0$).  For example, the disordered couplings drive to a low temperature CSG phase at lower intensities of magnetic fields, as pointed by the white region delimited by the AT line in Figs. \ref{iso_entropic}(d)-(i). However, the magnetic field suppresses gradually the CSG phase, at the same time that it can also induce a CSG close to critical fields, as depicted in Figs. \ref{iso_entropic}(g)-(i) for intermediate disorders \cite{Schmidt_2019}. 
To avoid the 
replica symmetry  solution instability, in the following, our entropy analyzes focus only on the PM phase.

Figure \ref{iso_entropic} shows clearly that
the magnetic field strongly affect the entropy, leading it to lower values (yellow color). On the other hand, the temperature drives to higher values of entropy (red color). However, the competition introduced by the geometrical frustration plays an important role in the entropy. This competition leads to higher entropy at lower temperatures, where the magnetic field brings out a peculiar entropic behavior that depends on the cluster size  (see first line, panels (a), (b) and (c), of Fig. \ref{iso_entropic}). For instance, the isentropic curves show an interesting response to magnetic fields (see solid lines in Fig. \ref{iso_entropic}).

In the clean limit, several adiabatic curves converge to  critical fields. In this case, we found the maximum gradient of the isentropics for fields close to the critical ones. Therefore, near the critical fields, a significant temperature drop can be driven by an isentropic field removal, indicating an optimal range of external field to explore the magnetocaloric effect. This phenomenon is a consequence of the energy levels crossing of the geometrically frustrated Hamiltonian, which leads to a large degeneracy. In this context, a natural question concerns the role of quenched disorder in the system: How these magnetocaloric properties introduced by the intracluster geometrical frustration are affected by disorder. We address this issue, in the following.

Considering a weak disorder ($T_f/J_1=0.1$ at $h=0$), we can observe a change in the behavior of the isentropes (see Figs. \ref{iso_entropic}(d)-(f)). The main changes take place in the neighborhood of the critical fields of the clean limit. In particular, the isentropes become smoother and the gradients near the critical fields are diminished. In addition, the convergence of several adiabatics to the same critical field is suppressed. 
However, at higher temperatures,  the  isentropes of the weakly disordered system resemble those of the clean limit. The complex energy landscape introduced by weak disorder soften the energy crossing level presented in the clean limit, which is manifested specially in the ground-state. However, for intermediate temperatures, this disordered system still preserves magnetocaloric signatures of geometric frustration.  Nevertheless, this geometric frustration feature drops out as the disorder increase, as shown in the bottom panels of Fig. \ref{iso_entropic} for an intermediate disorder regime ($T_f/J_1=0.3$ at $h=0$). It is important to remark that these discussed effects appear at the PM phase with replica symmetry stability, above the CSG phase. Therefore, our theoretical calculations suggest that the isentropes of weakly disordered cluster magnets will be smoother when compared to the clean system.  We remark that results for a heptametallic gadolinium molecule present smoother isentropes when compared to the microscopic model calculations for the system \cite{sharples2014quantum}. Moreover, the difference between experimental and theoretical results is enhanced at lower temperatures. Our findings suggest that this discrepancy can be related to disorder effects on this molecular magnet, which are not incorporated within the model calculations presented in Ref. \cite{sharples2014quantum}.

Now, we discuss the magnetocaloric potential given by the isothermal entropy change $\Delta S$. For instance, 
Fig. \ref{delta_s} shows $\Delta S$ as a function of  temperature for different sceneries with disordered clusters.  The $-\Delta S$ exhibits a maximum value at a range of  temperature above the freezing one, in the PM phase, where the  replica symmetry  solution is stable. This maximum depends on the intensity of magnetic field variation $\Delta h$ (here, $\Delta h=h-0$ represents the difference between applied field $h$ and zero field). However, at a weak disorder regime, $-\Delta S$ does not increase monotonically with $\Delta h$ in the whole range of temperature (see Fig. \ref{delta_s}(a)).  At temperatures  near the maximum of $-\Delta S$, the increase of $\Delta h$ can cause a reduction of $-\Delta S$, mainly when $\Delta h$ approaches the values of critical fields (see, e.g., the data for $\Delta h/J_1 = 1.0$ and $\Delta h/J_1 = 2.0$). 
It means that an enhancement in the MCE is not always prompt by an increase of $\Delta h/J_1$ in  GF systems with weak disorder. 
Furthermore, one can observe in  Fig. \ref{delta_s}(a) that a small intensity of magnetic field ($\Delta h/J_1=1.0$) is able to induce a considerable isothermal entropy change.  For instance, the $-\Delta S$ maximum is around 0.24 at $\Delta h/J_1=1.0$, increasing to 0.54 at higher field $\Delta h/J_1=5.0$ when $n_s=6$. Therefore, in the following, results for low field variations are explored in order to optimize the contribution coming from the GF.

 Fig. \ref{delta_s}(b) shows the dependence of the isothermal entropy change with the disorder intensity at low field variation. The increase of disorder leads $T_f/J_1$ to higher temperatures, approaching it to the temperature ($T^*$) where the maximum of $-\Delta S$ occurs. In addition, the maximum of $-\Delta S$ decreases as the disorder enhance, mainly when the strength of disorder becomes intermediate and high ($T_f/J_1\geq 0.3$).  

The cluster size can also affect the magnetocaloric potential, as illustrated in Fig. \ref{delta_s}(c), where the increase of clusters size amplifies $\Delta S$ at weak fields variations. This result is robust in a weak strength of disorder, as can be seen by comparing panels (c) and (d) of Fig.  \ref{delta_s} for $T_f/J_1=0.1$ and 0.2, respectively. 

Our findings could be compared to the results of the system R$_2$Ni$_1$Si$_3$, \cite{PhysRevB.94.104414,Pakhira2018} where the R stands for a rare earth element. 
Such system has a P6/mmm structure, with the R elements forming a triangular lattice. 
The presence of antiferromagnetic ordering suggests certain geometric frustration to this R$_2$Ni$_1$Si$_3$ system. Experimental results to the compound Nd$_2$Ni$_1$Si$_3$, Gd$_2$Ni$_1$Si$_3$ and  Er$_2$Ni$_1$Si$_3$ indicate a small Curie-Weiss temperature of about 1.2 K, 14.3 K and 0.8 K, which could  hint ferromagnetic correlations due to the presence of disorder in these compounds. 
The observation of substantial freezing temperatures of 2.85 K, 16.4 K and 3 K, for the Nd-, Gd- and Er-Ni$_1$Si$_3$ compounds, respectively, are indication of the median- to high-disorder regime.
In all three cases, the maximum value of $-\Delta S$ versus temperature occurs just above the freezing temperature, and showing an asymmetrical profile, as predicted here. The values of $-\Delta S$ and relative cooling power found for these frustrated-disordered magnets are highly promising for low temperature refrigeration. 
Usually, frustrated systems with CSG phase present $T_f$ below 100 K, so this should the range of applicability of the present model. 
Our results also point out mechanisms for further improvements in such materials. For instance, 
disorder could stabilize the cluster-glass at high temperatures in detriment of MCE, so $\Delta S$ could be even enhanced by reducing the disorder level, putting the R$_2$Ni$_1$Si$_3$ in a weak disorder regime. Here, we have shown that $\Delta S$ could be enhanced by a factor 2 upon reduction of disorder level.

The maximum value of the isothermal entropy change $-\Delta S_{max}$ as a function of $\Delta h$ is exhibited in Fig.  \ref{max_entropy} for different  levels of quenched disorder and several cluster sizes.
For weak disorders, a non-monotonic behavior of $-\Delta S_{max}$ with $\Delta h$ occurs, in which local maxima appear displaced from the critical fields. These local maxima of $-\Delta S_{max}$ increase with the intensity of $\Delta h$.  However, close to critical fields, the energy level crossing
associated with the magnetization jumps can enhance the degeneracy at weak disorder regime, leading to higher entropy at the applied field and, consequently, decreasing the isothermal entropy change.
It is important to remark that the structure of local maxima depends on the cluster size. In other words, this structure is a consequence of the presence of the magnetic plateaus of the geometrically frustrated clusters (see Fig. \ref{entropy_clean}), which can persist at weak quenched disorders. However, the increase of disorder can smooth the magnetic plateaus, eliminating the effects of GF on the entropy behavior.  
In addition, for a wide range of $\Delta h$ the $-\Delta S_{max}$ becomes lower as disorder increases. 
We also note that when the quenched disorder turns intense,  the $-\Delta S_{max}$ gets a monotonic behavior with $\Delta h$, and the local maxima disappear.  
To summarize, in this case, one can identify two different regimes: for weak disorders,  the geometric frustration of clusters is relevant to the thermal behavior, leading to high values of $-\Delta S$ in low magnetic fields; for strong disorders, the intercluster interactions dominate the scenery, avoiding the GF effects and requiring higher magnetic fields to get a large  $-\Delta S$.
 
It is interesting to point out, from the technological perspective, that due to the magnetization plateaus, the entropy rapidly reduces by applying a small magnetic field (up to 0.3 $h/J_1$) and
remains constants up to the critical fields. 
It means that high relative cooling power could be
obtained even in the presence of small magnetic fields. The amount of $-\Delta S$ which could be obtained by using low magnetic fields increases by raising the cluster-glass size or by reducing the disorder level. 
At high temperatures otherwise, the system behaves like a usual paramagnet, and much higher magnetic field will be needed to obtain the same relative cooling power.

\section{Conclusion}

We study the magnetocaloric effect of geometric frustrated systems in the presence of clusters with quenched disordered interactions. The model considers a network of clusters with random interactions, where the clusters are composed by Ising spin triangular structures with antiferromagnetic interactions. The replica method is used to decouple clusters, which are then treated by exact enumeration. Phase diagrams and entropy behavior are analyzed for different cluster sizes and strength of disorder in order to obtain the magnetocaloric potential for several sceneries.

We identify two regimes concerning the relation between geometric frustration and disorder: a weak disorder one, which can lead to a CSG phase at lower temperature and still preserve some geometric frustration effects above the freezing temperature; and strong disorder regimes, where a glassy order occurs without geometric frustration traces. In general, the isentropic curves are strongly affected by disorder at low temperature. 
In the clean system, they converge to critical fields that depend on the cluster size, but a weak disorder prevents this convergence at zero temperature.
However, only intermediate and strong disorder regimes affect the isentropic behavior at higher temperatures. 
It means that the geometric frustration can still drive the entropy behavior at a relevant range of finite temperature at weak disorder.  
In fact, the isothermal entropy change reaches a maximum at temperatures above the freezing one, within the paramagnetic phase. 
This maximum value depends on the disorder strength. However, for weak disorder regimes, the contribution of the geometric frustration has been shown relevant to enhance the magnetocaloric potential in low intensities of external magnetic fields. Furthermore, $-\Delta S$ increases with the cluster size.  In this case, the weak disorder, that drives to a network of clusters, can help to increase the magnetocaloric potential.
The ultimate mechanism to determine such increase is, at temperatures $T>T_f$, the weak disordered interaction activates the cluster magnetic moment, picking among the manifold of degenerate spin configurations due to the geometric frustration, the one whose total spin of the cluster is maximized. Under the action of a magnetic field, the additional degree of freedom introduced by the cluster network produces huge releases of entropy.

The present theory can also provide a useful framework to investigate the enhancement of MCE in 
new cluster systems with triangular structure \cite{PhysRevB.103.214427} and molecular magnets \cite{Brown_2016} with weak randomness and strong frustration. 
Our findings suggest that weak disorder leads to changes on the magnetocaloric effect of cluster magnetic systems only at very low temperatures. Intermediary levels of disorder, however, can lead to important effects at higher temperatures. In our opinion, the present disordered cluster model can be used as a first approximation to incorporate disorder effects on molecular magnets and other nanostructures.

\section*{Acknowledgments}
FMZ, RM, and SGM acknowledge the support from CNPq/Brazil. MS acknowledges the support of Funda\c{c}\~ao de Amparo \`a Pesquisa do Estado do Rio Grande do Sul (Fapergs). 
The authors are in debt with Luis F. Barquin for the critical reading and useful suggestions.

\section*{References}
\bibliographystyle{iopart-num}
\bibliography{references}
\end{document}